# A Predator Prey Approach to Diversity Based Defenses in Heterogeneous Networks

Sean P. Gorman*, Rajendra G. Kulkarni,
Laurie A. Schintler, Ph.D., and Roger R. Stough, Ph.D.

School of Public Policy, George Mason University, Fairfax, Virginia 2200, U.S.A.
*Corresponding author
e-mail: sgorman1@gmu.edu

## ABSTRACT

In light of the rise of malicious attacks on the Internet and the various networks and applications attached to it, new approaches towards modeling predatory activity in networks is called for. Past research has simulated networks assuming that all vertices are homogenously susceptible to attack or infection. Often times in real world networks only subsets of vertices are susceptible attack or infection in a heterogeneous population of vertices. One approach to examining a heterogeneous network susceptible to attack is modeling cyberspace as a predator prey landscape. If each type of vulnerable device is considered a heterogeneous species what level of species diversification is needed to keep a malicious attack from a causing a catastrophic failure to the entire network. This paper explores the predator prey analogy for the Internet and presents findings on how different levels of species diversification effects network resilience. The paper will also discuss the connection between diversification, competition, anti-trust, and national security.

## INTRODUCTION

The "next big thing" in cyberspace is no longer glitzy new technologies, but malicious agents that have been wreaking havoc on the Internet, such as viruses, worms, Trojans and denial of service attacks (Economist 2003). While the problem of malicious agents has been widely studied relatively there is significantly less research that has endeavored to look at the problem from a theoretical or policy perspective. This paper will look at the problem utilizing the theoretical under pinning of predator prey models from biological science as well as recent research in network complexity and mechanics.

Recent research has found that several critical technological networks are scale free structures with power law connectivity distributions, such as the Internet at the autonomous system level and the router level (Faloutsos et al 1999), the World Wide Web (Barabasi et al 1999, Huberman and Adamic 1999), and physical SDH telecommunications networks (Spencer and Sacks 2003). Studies have examined the vulnerability of scale free networks finding that they are resilient to random attacks, but highly susceptible to targeted attacks (Albert et al 2000). Researchers have also examined the immunization of scale free networks including the Internet, finding that targeted immunization strategies work well, while random strategies fail to eradicate a virus below an epidemic threshold (Pastor-Satorras and Vespignani 2001, Pastor-Satorras and Vespignani 2002, Dezsos and Barabasi 2002). These studies, though, have analyzed scale free networks assuming that all vertices are homogenously susceptible to attack or infection. Often times in real world networks only subsets of vertices are susceptible to attack or infection in a heterogeneous population of vertices. One example of such a scenario is Internet worms, which are designed to attack/exploit only specific operating systems or platforms. One approach to examining a heterogeneous network susceptibility to attack is as a predator prey system. If each type of vulnerable device is considered a heterogeneous species, what level of species diversification is needed to keep a malicious attack from a causing a catastrophic failure to the entire network? To answer this question first a predator prey typology for the Internet will be introduced, followed by a review of predator prey research in other fields, and finally a methodology for applying a predator prey model to the Internet with preliminary results and conclusions.

**A PREDATOR PREY TYPOLOGY**

The Internet and the wide array of networks and applications interconnected to it are plagued by a variety of malicious agents, ranging from viruses to worms to denial of service attacks. In analogy, the Internet can be viewed as an ecosystem that connects a wide variety of habitats such as routers, servers, operating systems (OS), etc. These habitats can further be delineated by the species that reside in them, for example a router habitat could comprise of Cisco, Juniper, Lucent, and Nortel species of routers. Using the Internet as an ecosystem analogy the malicious attacks outlined above (worms, viruses, etc.) can be seen as predators and their interaction with ecosystem (servers, routers, OS's) as a predator prey relationship. A sample typology of a predator prey ecosystem with examples can be viewed in figure 1.

Figure 1. Predator Prey Sample Typology for the Internet

- *Ecosystem* - The Internet
    - *Habitats*
        - Routers
            - *Species*
                - Cisco
                - Juniper
                - Lucent
                - Nortel
        - Servers
            - *Species*
                - Microsoft - IIS
                - Apache
                - Unix
                - BSD
        - Operating Systems
            - *Species*
                - Microsoft – Windows
                - Macintosh
                - Unix
                - Linux
        - Email Applications
            - Species
                - Microsoft – Exchange
                - Web based email
                    - Hotmail
                    - Yahoo

- AOL
- Netscape

This paper will attempt examine how a predator prey analogy can be modified to the Internet in order to gain insight into the dynamics and security of the system. To do so a quick review of the fundamental ideas behind predator prey models will be covered, followed by how these models can be adapted to the Internet, how these models can provide insight into improved Internet security and stability, and the policy implications of these findings.

*Lotka-Volterra Approach*

The literature around the ideas of predator prey models is well developed. The mathematical beginning to predator prey models was the competition models independently created by Lotka (1925) and Volterra (1926). The Lotka-Volterra models explained oscillations in populations between competing species. In simple terms when there is a large number of prey the environment is good for predators and their population increases. This predator population increases until there are not enough prey to sustain the population and then the predator population decreases. The decrease in predators then leads to an increase in prey since there are fewer predators consuming them. This oscillating relationship between predators and prey can be mathematically expressed as:

$$\frac{dH}{dt} = H(a - \boldsymbol{a}P)$$

$$\frac{dP}{dt} = P(-b + \boldsymbol{b}H)$$

- $H(t)$ and $P(t)$ represent the magnitude of prey and predator populations respectively.
- $a$ is the growth rate of the prey population in the absence of predators.
- $b$ is the rate at which a predator population will decrease without sufficient amount of prey to feed upon.
- $\boldsymbol{a}$ and $\boldsymbol{b}$ measure the negative impact of predators on prey and the positive impact of prey on predators, respectively.

(Waage and Mills 1992)

The basic linear Lotka-Volterra model has been modified to examine non-linear relationships (Paine 1966, Paine and Hanski 1998), cooperation among competitors (Zhang 2003), enhanced stability (Gonzales-Olivares and Ramos-Jiliberto 2003), and geographic effects (Tainaka 2003) to name just a few of the models and researchers' extensions. The Lotka-Volterra model has also been used to explain relationships other than the predator prey dynamics in biology, including such diverse disciplines as economics, chemistry, physics, mathematics, geography, and demography. This research proposes applying the predator prey model to a new area, complex networks, specifically the Internet and the wide variety of malicious attacks it faces.

Previous work points towards applying predator prey models to the Internet as being a fruitful endeavor. Lopez et al (2003) found that Lotka-Volterra equations could reproduce the competitive structure of a complex network of web sites. Further, Solomon (1997) found that a generalized Lotka-Volterra equation produces a power law probability distribution, specifically with the distribution of wealth in a simulated market. Both papers indicate that Lotak-Volterra models could produce the competitive forces and resulting network structure seen in scale free networks. While this paper will not address this issue directly the possible connection is an important one as a predator prey model for the Internet is developed, since it has been widely found to be a scale free network at several levels (Falustosos et al 1999, Albert et al 1999).

*Predator-Prey for the Internet*

The first step in developing a predator prey model for the Internet is to outline how electronic predators and prey differ from what is found in the natural world. To begin this one needs to define what a predator and prey will be in terms of the Internet. For the purposes of this discussion predators are any malicious agents that can cause damage to systems, applications and software connected to the Internet, and the prey are the same systems, application and software that can be damaged. Predators come in a variety of species, some self propagate like worms and viruses, and others like denial of service attacks and Trojans are static, and there are hybrids of the two. Depending on the behavior of the predatory species malicious attacks must be modeled differently in relation to how they interact with prey. For the purposes of example Internet worms will be illustrated. A worm, depending on how it is programmed, attacks specific species across the Internet. For example the SQL slammer worm of January 2003 was programmed to attack Microsoft SQL server machines. The worm only attacks one specific species and propagates itself through the attacked machines. This method of propagation differs from what is seen in most predator prey models since prey are basically turned into a predator to attack other prey and in turn change them to predators. While this behavior is not often seen in animals it is analogous to what happens with cancer at the cellular level. A cancer infects a cell turning it into a cancer cell, which further propagates the cancer to other cells. As a result predator and prey populations fluctuate on a quasi one to one basis, the predator population can only increase if a prey is converted to a predator. The prey population on the other hand can increase from the removal of a predator or the production of additional prey connected to the Internet.

A second key differentiator is the growth rates of predator and prey populations in the ecosystem. Predators can have variable rates of growth. A worm, for instance, can have an explosive rate of growth spreading across the entire ecosystem in as little as ten minutes (Moore et al 2003). While the growth of worms is often highly rapid, their population decline can also happen quickly. Worms most often affect network habitats that can evolve defense strategies very rapidly that block the worm through filters and/or patch the vulnerabilities in the prey, which allow the worm to propagate. As a result worms can have a quick and often devastating effect on a prey population, but that high level of risk has led to the evolution of defenses that can mitigate the longevity of these predators. Viruses on the other hand tend to have a slower growth rate targeting end user habitats typically through email applications (species). While viruses tend to propagate at a slower rate than worms their prey are typically less sophisticated in their defenses and slower to evolve. Defense strategies for viruses typically come from anti-virus software, but implementation of these defenses are many times left to end-users and are not often coordinated or evolve rapidly over time. As a result viruses often persist with large populations for extended periods of time, often never completely disappearing until their prey becomes extinct.

In may real world scenarios there is no warning for cyber attacks or what vulnerability may be exploited, thus there is no premeditated defense of the attack. This leaves the difficult scenario of how does one defend a network when one does not know what the attack will be. The predator prey analogy offers a means of defense through species diversification. Since most predators can only effect one species at a time, the diversification of species in a network could provide a natural defense against attack. This research will investigate what is the minimal amount of diversification needed to prevent any one predator from having a catastrophic effect on an ecosystem. A predator prey approach can be further developed to investigate how prey evolves defenses to predator attacks and how predators in turn evolve to thwart these defenses. It may be possible to simulate the evolutionary tracks of the predator prey relationship to discover how security and defense might best be maximized and predator populations minimized.

**PREDATOR-PREY METHODOLOGY**

The methodology of the approach will be to start with a simulated scale free network with 12,000 vertices, in which all vertices a homogenous. Next, 1% of those vertices will be randomly changed into a different species and the network will then be heterogeneous. At this point a predator will be introduced to the population that only preys upon the new species, at this step only 1% of the population. After the predator has spread through the susceptible prey population the number of disconnected vertices and the number of vertices with only one connection will be calculated. Next, 2% of the vertices will be randomly seeded as the new prey population and the process will be repeated. This process will be duplicated at 1% intervals until all the vertices in the network consist of the new species. The results for this procedure are illustrated in figure2.

Figure 2 – Number of Disconnected Vertices versus Size of Susceptible Prey Population with Random Seeding

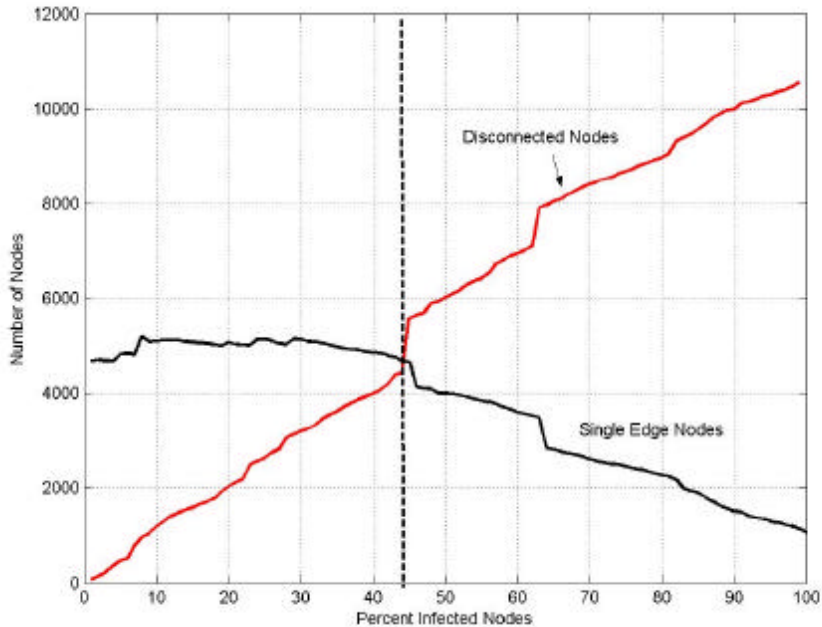

The number of disconnected vertices in the network increases linearly until 43% at which time there is an almost vertical jump. The large jump in disconnected vertices coincides with the crossover with number of single edge vertices that experiences a complimentary steep drop. The combination of steep numerical shift and cross over is indicative of a catastrophic failure in the network. When 43% of the population becomes any single one species a single predator can cause traumatic damage to the total network. At the 43% point there are more disconnected vertices than single edge vertices and large parts of the networks can longer communicate with each other. The same break point can be seen when the number of disconnected edges is plotted as seen in figure 3.

Figure 3: Number of Disconnected Edges versus Size of Susceptible Prey Population for Random Seeding

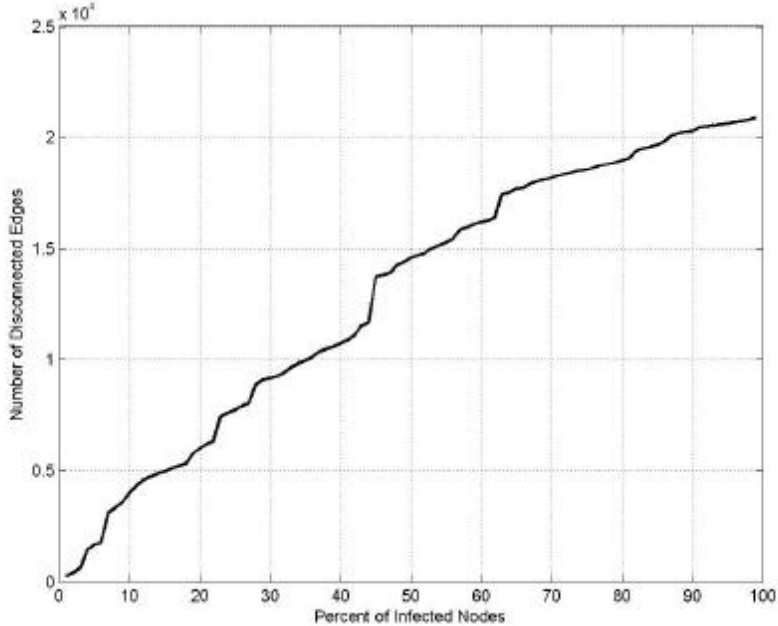

Another way to view the degradation of the network is by looking at the data by segregating vertices by their number of edges. This was done by constructing a histogram with midpoints at 25, 50, 100, 250, 500 edges per vertex. The output of this approach can be seen in figure 4.

Figure 3. Vertex Failures Segregated by Number of Edges for Random Seeding

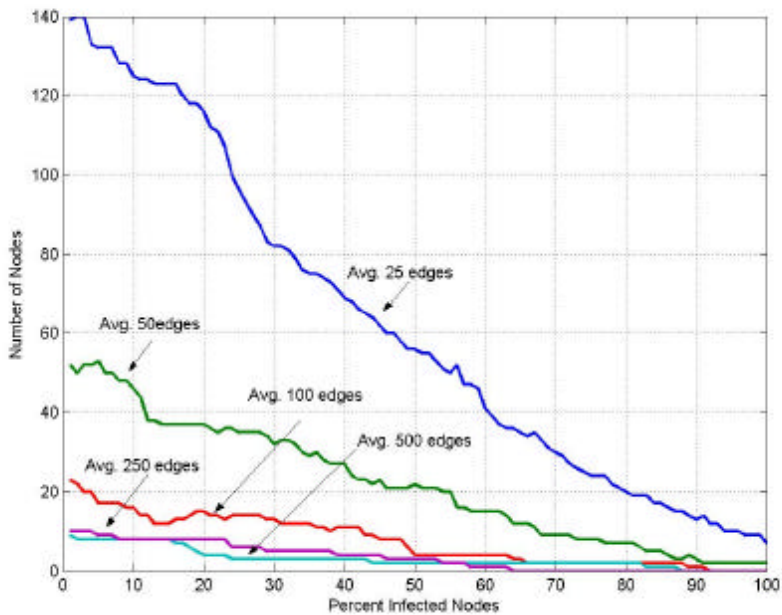

The histogram approach illustrates that 25 average edges decreases most rapidly overall, while vertices with an average of 500 edges experiences a sharp drop after 20% of vertices are infected. The effect on vertices with large number of edges is particularly important because of the heavy dependence on the vertices to connect the poorly connected vertices that constitute the majority of the network (Albert and Barabasi 2002). The results do seem to indicate that vertices with higher connectivity are more resilient to increasingly susceptible prey populations.

Another variation on this procedure is to use a targeted seeding of the prey population instead of a random seeding. Specifically starting with the most connected vertex as the first member of the new species, and the second most connected, third most connected, and so on until the least connected vertex is turned into the new species. The number of disconnected vertices and the number of single edge vertices is calculated for every 1% interval. The results of this targeted approach are shown in figure 3.

Figure 6 – Number of Disconnected Vertices versus Size of Susceptible Prey Population with Targeted Seeding

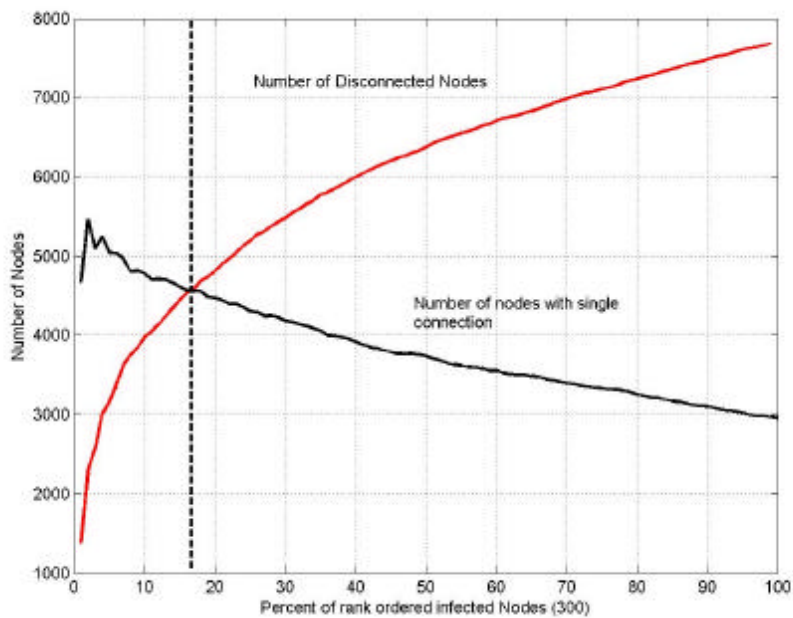

The results of the targeted approach illustrate a far more rapid degradation of the network, an exponential increase in the number of the disconnected vertices versus the linear degradation seen in the random seeding. Also the crossover point with the number of single edge vertices occurs far earlier around 17%, coinciding with the inflection point of the disconnected vertices curve. While the simulation is not realistic it does demonstrate the point, that when a species clusters amongst the more connected vertices in a network it dramatically increases the vulnerability of that network. The corollary to this result would be the more diversity in the core of the network the more robust the network. Viewing the vertex degree histogram plot used with the random seeding approach reinforces this corollary. The same distribution of average connectivity at 25, 50, 100, 250, 500 midpoints was again used, and is displayed in figure 6.

Figure 6. Vertex Failures Segregated by Number of Edges for Targeted Seeding

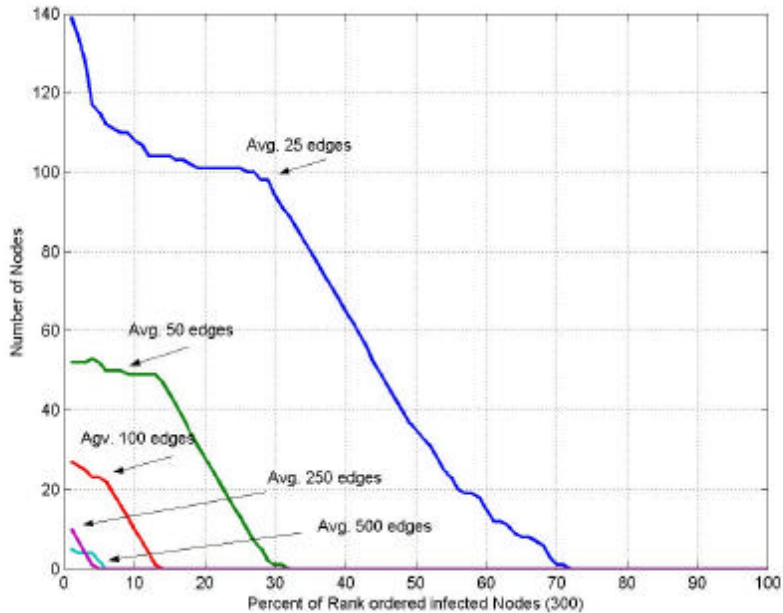

The shape of disconnected vertex arc for the 25, 50, and 100 edges are all very similar, while the 250 and 500 arcs down to zero almost immediately. The rapid failures of highly connected vertices are directly correlated with the targeted seeding strategy. In the targeted seeding strategy the most connected vertices are made susceptible first, so their rapid failure would be expected. The rapid failure of the network in general is linked to the failure of the most connected nodes first. This is reinforced when the number of disconnected edges is presented, seen in figure 7.

Figure 7. Number of Disconnected Edges versus Size of Susceptible Prey Population for Targeted Seeding

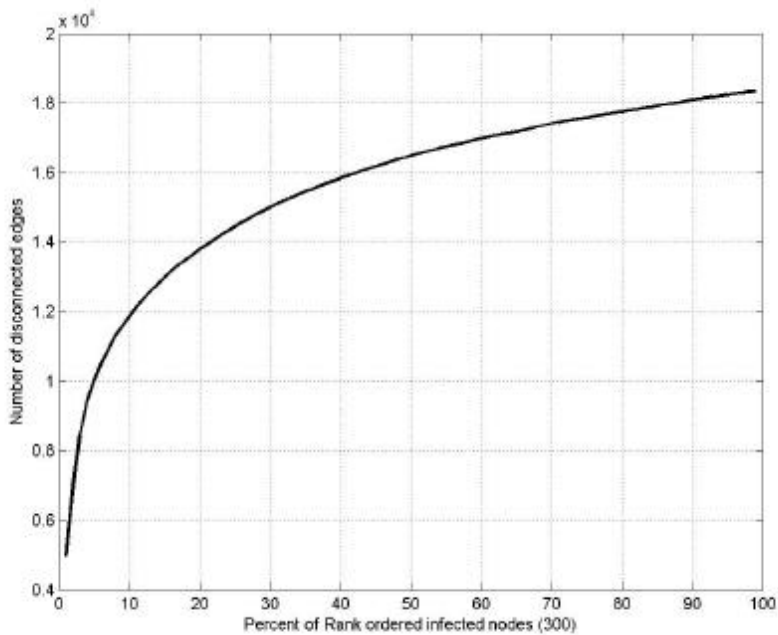

The targeted seeding results in an exponential loss of edges in the network, as opposed to the linear loss seen in the random scenario. The finding reinforces that the greater homogeneity in the highly connected core the more vulnerable the network.

**CONCLUSIONS**

The findings of this preliminary research illustrate the security problems of reliance on a single network application (species). The results should not been viewed as hard numbers, by which decisions could be based. The approach is useful when questions arise if the market share of a particular company in a networked environment is causing security vulnerability on a macro scale. Results could vary widely by network topology, type of attack, and vulnerability. The particular network used for these simulations is indicative of a large number of critical global infrastructures, including the Internet at the router and autonomous system level and the World Wide Web. The similarity of the network tested in this research to those critical infrastructures raises the question if monopoly or quasi monopoly conditions can contribute to national security vulnerabilities. Further, if such vulnerabilities bear out is antitrust policy an appropriate response. Antitrust regulation falls under the Sherman Act and specifically, the Supreme Court said in its *[Professional Engineers](#) Case*, 435 U.s. 679, 695 (1978):

> The Sherman Act reflects a legislative judgment that ultimately competition will produce not only lower prices, but also better goods and services. "The heart of our national economic policy long has been faith in the value of competition." Standard Oil Co. v. FTC, 340 U.S. 231, 248. The assumption that competition is the best method of allocating resources in a free market recognizes that all elements of a bargain - quality, service, safety, and durability - and not just the immediate cost, are favorably affected by the free opportunity to select among alternative offers.

In this case the safety clause would be the part of legislation that specifically applies to network-based applications. When there is a lack of competition and a resulting lack of diversity this research points to a pronounced negative safety externality. The interdependent and interconnected nature of network-based applications exacerbates the safety issue by affecting constituents who do not use the product or application. In the case of this research those constituencies are represented by the disconnected vertices and were the non-susceptible prey in each simulation.

The problem of a growing lack of diversity in networks can be exacerbated by technology lock-in, "A customer experiences "lock-in" when switching costs exceed the potential incremental value of alternative suppliers' products over its current supplier's product (Lookabough and Sicker 2003)". Lock-in has been a common technology strategy in recent years directed towards growing market share and producing stable revenue growth (add cite). When a lock-in strategy aggregates across any one platform the result can be a non-optimal market creating interdependent security vulnerabilities and negative externalities. Lookabough and Sicker (2003) state that, "security induced lock-in has resulted in convergence to a stable equilibrium that is not the globally optimal one" and point towards antitrust policy as one possible policy remedy.

Utilizing national security as a precedent to enforce antitrust policy to promote competition has historical grounding. During World War II "national security was a potent weapon in the battle for both public investment and antitrust enforcement (Hart 1998)." Interestingly US policy to date has gone in the other direction with their legislative efforts. The Critical Infrastructure Security Act of 2001 (S 1456 IS) states that, "antitrust laws inhibit some companies from partnering with other industry members, including competitors, to develop cooperative infrastructure security strategies." While closer cooperation among industry actors is vital it is important not to shelve antitrust policy options in the process. This research highlights the benefits of diversity in increasing the robustness of a network and there is a direct link between competition and increased diversity. Antitrust is one policy tool that has been successfully used by the US government to promote national security in the past and it is possible that is could serve that purpose again.

This is just one application of using a predator prey model to simulate the Internet. The model is still rudimentary and needs further exploration, but the initial findings appear promising with interesting implications. Future directions include examining real world networks, market shares, and vulnerabilities. Also a deeper investigation of applying the Lotka-Volterra equation to real world virus and worm propagations could yield helpful understandings of how to minimize malicious attack damage in the future.

**WORKS CITED**

Albert R, Jeong H, Barabasi A, 1999, "The diameter of the World Wide Web" *Nature* 401: 130-131

Albert R., Jeong H., and Barabási A.L., 2000, Attack and error tolerance in complex networks. *Nature* 406: 378.

Albert R, Barabási, A, 2002, "Statistical mechanics of complex networks" *Reviews of Modern Physics* 74: 47-97

Batty, M, 1997, "Virtual Geography", *Futures*, Vol. 29, No. 4/5, pages 337-352.

Dezsos, Z., Barabasi, A.L., 2002, Halting viruses in scale-free networks, Physical Review E 65: 055103 (R).

Faloutsos C, Faloutsos P, Faloutsos M, 1999, "On power-law relationships of the Internet Topology" *Computer Communication Review* 29: 251-260

Gonzales-Olivares, E, Ramos-Jiliberto, R, Forthcoming, Dynamic consequences of prey refuges in a simple model system: More prey, fewer predators, enhanced stability" *Ecological Modeling*

Hart, DM, 1998, *Forged Consensus: Science, Technology, and Economic Policy in the United States, 1921-1953*, Princeton, N.J.: Princeton University Press

Huberman B, Adamic L, 1999, "Growth dynamics of the World Wide Web" *Nature* 401:131-134

Lookabough, T, Sicker, DC, 2003, "Security and Lock-in" Economics and Information Security Workshop, University of Maryland – http://www.cpppe.umd.edu/rhsmith3/papers/Final_session8_lookabaugh.sicker.pdf

Lotka, AJ, 1925, *Elements of Physiological Biology* Dover Publications: New York.

Moore, D., Paxson, V., Savage, S., Colleen, S., Staniford, S., and Weaver, N, 2003, *The spread of the Sapphire/Slammer worm.* CAIDA - http://www.caida.org/outreach/papers/2003/sapphire/sapphire.html

Paine, RT, 1966, "Food web complexity and species diversification" *American Naturalist* 100: 65-75

Pastor-Satorras, R., and Vespignani, A., 2002, Immunization of complex networks, Physical Review E 65: 036104-1.

Pastor-Satorras, R., Vespignani, A., 2001, Epidemic dynamics and endemic states in complex networks, Physical Review E 63: 066117.

Solomon, S, 1997, "Stochastic Lotka-Volterrs systems of competing auto catalytic agents lead generally to truncated Pareto power wealth distributions, truncated Levy distribution of market returns, clustered volatility, booms and crashes" in Eds. Burgess, AN, Moody, JE, *Computational Finance* Kluwer Academic Publishers.

Spencer J. and Sacks L., "On Power-Laws in SDH Transport Networks" IEEE ICC 2003, May 2003, Anchorage, Alaska, USA

Tainaka, K, 2003, "Pertubation expansion and optimized death rate in a lattice ecosystem" *Ecological Modeling* 163: 73-85

Volterra, V, 1926, "Fluctuations in the abundance of a species considered mathematically" *Nature* 118: 558-560

Waage, JK, Mills, NJ, 1992, "Biological Control" in (Eds.) Crawley, MJ, *Natural Enemies: The Population Biology of Predators, Parasites and Diseases* Blackwell Scientific Publications: Oxford

Zhang, Z, 2003, "Mutualism or cooperation among competitors promotes coexistence and competitive ability" *Ecological Modeling* 164: 271-282